\documentclass[aps,prl,twocolumn,showpacs,superscriptaddress,floatfix]{revtex4}
\usepackage{graphicx,amsmath,amssymb,color}
\begin{document}

\title{Fractional Magnetization Plateaus and Magnetic Order in the Shastry
Sutherland Magnet TmB$_4$}

\author{K.~Siemensmeyer}
\author{E.~Wulf}
\affiliation{Hahn Meitner Institut Berlin, Glienicker Str. 100, D 14109 Berlin, Germany}
\author{H.-J. Mikeska}
\affiliation{Institut f\"ur Theoretische Physik, Universit\"at Hannover, 30167 Hannover, Germany}
\author{K.~Flachbart}
\author{S.~Gab\'{a}ni}
\author{S.~Ma\v{t}a\v{s}}
\author{P.~Priputen}
\affiliation{Institute of Experimental Physics, Slovak Academy of Science, SK 04001 Kosice, Slovakia}
\author{A.~Efdokimova}
\author{N.~Shitsevalova}
\affiliation{Institute for Problems of Material Science, Ukraine Academy of Science, UA 03680 Kiev, Ukraine}

\date{\today}

\begin{abstract}

We investigate the phase diagram of TmB$_4$, an Ising magnet on a frustrated
Shastry-Sutherland  lattice by neutron diffraction and magnetization
experiments. At low temperature we find N\'{e}el order at low field,
ferrimagnetic order at high field and an intermediate phase with magnetization
plateaus at fractional values $M/M_{\rm sat} = 1/7, 1/8, 1/9 \dots $ and
spatial stripe structures. Using an effective $S=1/2$ model and its equivalent
two-dimensional (2D) fermion gas we suggest that the magnetic properties of TmB$_4$
are related to the fractional quantum Hall effect of a 2D electron gas.

\end{abstract}


\maketitle

The magnetic properties of quantum spins with antiferromagnetic (afm)
coupling on frustrating lattices have attracted widespread interest in recent
years due to the discovery of new types of complex quantum ground
states. Recent examples include the spin liquid dimer phase observed in
SrCu$_2$(BO$_3$)$_2$ [\onlinecite{Kag99}] (a 2D magnet
on the Shastry-Sutherland lattice (SSL) \cite{ShaS81}), supersolid phases on
triangular and kagome lattices \cite{MoeS01} and the spin ice state in
pyrochlore \cite{BraG01}.  The interest in such systems is in particular in
magnetization plateaus at fractional values of the saturation magnetization as
observed in SrCu$_2$(BO$_3$)$_2$ \cite{KodTH02} and in fractionalized
excitations as discussed for spin ice \cite{FulPS02}, in particular the
possible existence of a magnetic monopole \cite{CasMS08}. This documents 
that plateaus and fractional excitations which are well known 
in quantum Hall physics also emerge in quantum magnets.

A relation between magnetization plateaus in 2D quantum spin systems and the
quantum Hall effect was established by a mapping of 2D S=1/2 magnets to
the 2D electron gas \cite{JolMG02}, requiring an additional Chern-Simons
field, a fictitious nonlocal strong magnetic field. When this field is 
spatially averaged it combines with lattice
induced minigaps (known from the 'Hofstadter butterfly' \cite{Hof76}) to
induce fractional magnetization plateaus. In simulations using this approach
fractional magnetization plateaus down to values as small as $M_{\rm sat}=1/9$
have recently been reported \cite{Seb07} together with complex modulated
spatial structures of the corresponding ground states. For investigations of
these phenomena in SrCu$_2$(BO$_3$)$_2$, however, the very high critical field
in this material is a severe limit and complex spatial structures
have not been found experimentally
so far.

In the following we present the results of magnetization and 
neutron diffraction experiments for TmB$_4$,
another 2D magnet on a SSL and fully accessible to experiments up to
the saturation field. We suggest that this compound is an excellent
realization of a frustrated magnet on the SSL, we find the emergence of
fractionalized magnetization plateaus strikingly similar to the case of
SrCu$_2$(BO$_3$)$_2$ and we observe spatial structures similar to those
described in Ref. \onlinecite{Seb07}. We find it attractive to speculate, on
the basis of experimental results to be described in the following, that
TmB$_4$ represents a magnetic analogue to the fractional quantum Hall effect (FQHE).

TmB$_4$ belongs to a group of rare earth tetraborides (REB$_4$) that
crystallize in a tetragonal lattice (space group 127) \cite{EtoH85}. The rare
earth moment position can be mapped to a SSL with nearly equivalent bond
lengths between nearest neighbors. In contrast to SrCu$_2$(BO$_3$)$_2$,
however, (i) TmB$_4$ is a metal and (ii) the Tm$^{3+}$ moment is $J=6$ and
subject to strong crystal field (CF) effects. Thus the RKKY interaction is expected
to be important and the basic spin model is an effective S=1/2 model
close to the Ising limit rather than a Heisenberg model.

{\it Experimental:} Single crystals of TmB$_4$ were grown by an inductive,
crucible-free zone melting method. The residual resistivity ratio was
larger than 100, documenting the high sample quality. The neutron sample had
cylindrical shape with a diameter of $\approx 7 mm$ and a length of $\approx 20 mm$, 
isotope enriched $^{11}$B was used to reduce the absorption. For
magnetization experiments oriented samples were cut with approximate dimension
1x1x1 $mm^3$. Magnetization data were measured using a commercial magnetometer
\cite{PPMS}. For neutron diffraction experiments we have used the E10
diffractometer at the HMI \cite{instrument}. A wavelength
of $0.14~nm$ was obtained from a Ge (220) monochromator. The $\lambda/2$
contamination was removed by Sapphire- and PG-filters. 

\begin{figure}
\includegraphics[width=80mm]{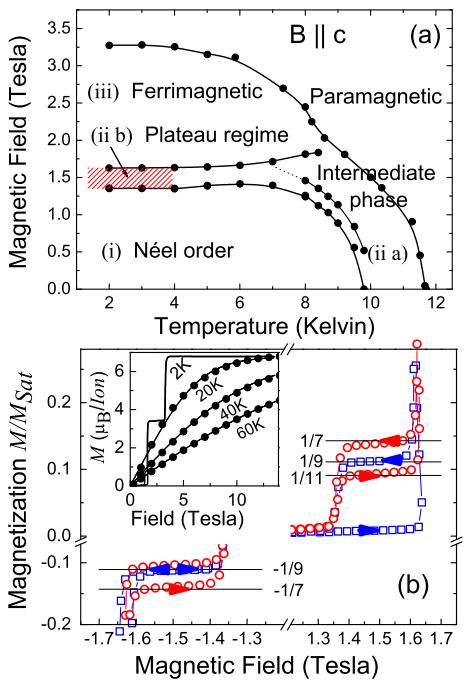}
\caption{(a) The phase diagram of TmB$_4$ as derived from
magnetization data for $B \ || \ (001)$ (full symbols). (b) Magnetization at
$T=2K$ (open squares) and $T=3K$ (open circles). The arrows indicate the
direction of the field change. The inset in (b) gives an overview towards high
field, lines are measured data, full circles are calculated using the high
temperature approximation. The magnetic field data are corrected for demagnetization.}
\label{fig:static}
\end{figure}

{\it Macroscopic properties:} TmB$_4$ shows a rich phase diagram as a function of field and temperature
(Fig.~\ref{fig:static}(a)) \cite{Yos06,Iga07}. In zero field magnetic order sets in at $T_N = 11.8 K$,
the afm low temperature N\'{e}el phase is stable below 9.8 K. Based on
magnetization experiments, at low temperature three phases are seen:
The 'high' field phase is ferrimagnetic with a plateau at $M/M_{\rm sat} = 1/2$. In the intermediate phase,
magnetization plateaus with $M/M_{\rm sat} = 1/7, 1/8, 1/9 \dots$ are seen at
temperatures below 4K. Their observation is subject to hysteresis. 
After zero field cooling plateaus are observed only when the ferrimagnetic phase has been reached once, 
but then independent on the direction of the magnetic field changes and the
field direction. The value of plateau magnetization varies between different
runs: although the energy difference between plateaus may be very small, the
time required for a significant spatial rearrangement of moments likely exceeds the time available experimentally.
We note that close to $T_N$ the intermediate phase splits into more complex phases
as identified in specific heat and resistivity measurements
\cite{Gab08,Iga07}, but with little signature in neutron diffraction and
magnetization experiments.

The single ion magnetic ground state is derived from the magnetic specific
heat \cite{Gab08} combined with magnetization data: The specific heat can be
analyzed in terms of a CF Hamiltonian $H = -g \mu_B B S_z + D
S_z^2$ where $D<0$ is the CF anisotropy. Towards high temperature, the 
entropy reaches the value $R \ln2$ which suggests an
effective doublet ground state for the noninteracting Tm$^{3+}$~ion. At low
temperature, for magnetic field parallel to the $c-$axis, the magnetization
reaches the saturation moment of Tm$^{3+}$ for $B > 5T$.  Therefore, the local
single ion ground state is the doublet $M_J = \pm 6$, implying that the
degeneracy of the $J = 6$ multiplet is completely lifted by the CF. The
splitting between ground state and first excited state can be estimated from
the Schottky anomaly \cite{Gab08}, yielding $D = 11K$. Consequently, the first
excited doublet with $M_J = \pm 5$ is at $\approx 100K$ and strong Ising
anisotropy is expected. However, transverse exchange will have to be added to
the dominant Ising coupling to improve the agreement of magnetization and
specific heat data to results for an ideal Ising model.

\begin{figure}
\includegraphics[width=80mm]{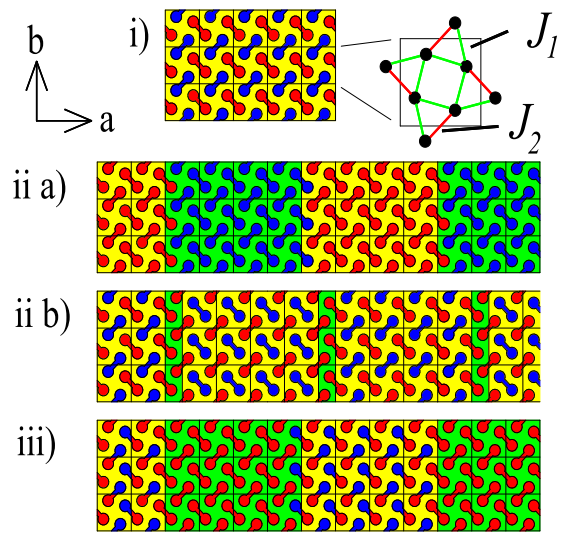}
\caption{The magnetic structure of Ising spins in TmB$_4$ in the tetragonal plane. Only 
domains along the a-direction are shown. Red(blue) circles indicate up(down) moments,  
the different colors indicate the periodicity of the various stripe structures.
Labels (i) to (iii) refer to the phase labels as used in
Fig.~\ref{fig:static}(a). In (i) the position of the Tm$^{3+}$ ions in the tetragonal plane is 
expanded with the SS bonds marked in red. Note the almost identical length of exchange paths $J_1$ and $J_2$ .} 
\label{fig:structure}
\end{figure}

{\it Magnetic structure:} The low temperature - zero field structure is the
N\'{e}el antiferromagnet (Fig.~\ref{fig:structure}(i)) with moments along the
(001) direction, implying ferromagnetic (fm) correlations along the
Shastry-Sutherland (SS) bonds. This follows from the observation of (100) and (300) reflections and
the absence of af intensity along $(00L)$.

The diffraction data for the high temperature phase
(Fig.~\ref{fig:structure}(ii a)) close to $T_N$ at $B =0$ agree well with an
amplitude modulated structure of fm stripes parallel to the a-axis,
separated by a $\pi$ domain wall. Any 'checkerboard' arrangements in diagonal
blocks can be excluded.

\begin{figure}
\includegraphics[width=80mm]{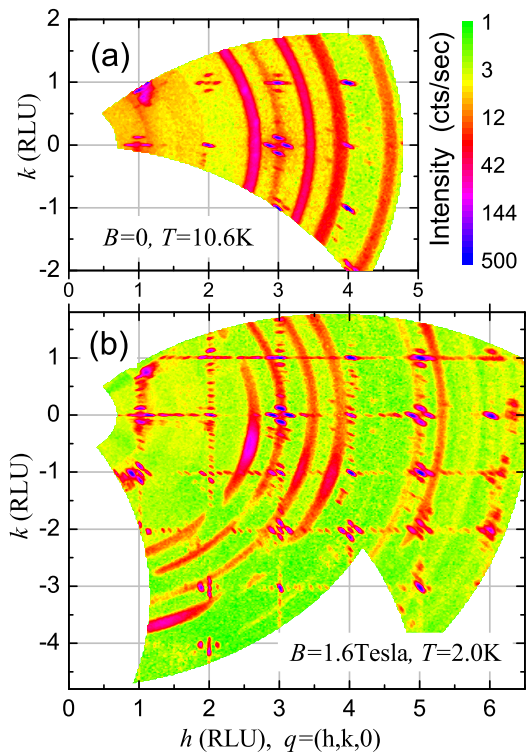}
\caption{(a) and (b) show the neutron diffraction pattern in
the tetragonal plane for the intermediate and the plateau phase,
respectively. The figures share the intensity color code given in (a). For the
indexing of the afm peaks see the text. The powder lines arise from aluminum in
the 'Orange' cryostat (a) and the 'VM3' magnet cryostat
\cite{instrument} used for experiments in a field (b).}
\label{fig:diffraction}
\end{figure}

For neutron investigations of the intermediate phase field cooling was
 used to avoid the hysteresis. This leads to the coexistence of magnetic
 structures with different unit cell size: Fig.~\ref{fig:diffraction}(b) shows
 a diffraction pattern with reflections that may be indexed by $(2m/7,0,0),
 (2m/9,0,0), (n/7,K,0) \dots$, where $m,n$ are integers. Lines of weak
 "harmonic" reflections result from stripe structures parallel to the $(H00)$
 or $(0K0)$ direction with a size 7 or 9 unit cells
 (Fig.~\ref{fig:structure}(ii b)), directly related to the magnitude of the
 observed magnetization plateaus with $M/M_{\rm_sat} = 1/7, 1/9$.  In
 addition, the afm reflections known from the zero field N\'{e}el phase have
 significant intensity, giving evidence for fm correlations along the
 SS bonds.

The ferrimagnetic high field phase shows again peaks indexed by 1/8,
indicating a large unit cell. Using the experimental result $M/M_{\rm sat} =
1/2$, it can be analyzed similarly to the intermediate phase. Here the fm
correlations along the SS diagonals are changed to afm for 1/2 of the large
magnetic unit cell (Fig.~\ref{fig:structure}(iii)) whereas the remaining
moments are just fm. This again results in a stripe structure, any diagonal
arrangement of the moments can be excluded.

To summarize the experimental results we conclude that the spin
interactions appear to be (i) antiferromagnetic, (ii) highly anisotropic and
(iii) governed by the Ising like CF anisotropy. They lead to stripe structures
and magnetization plateaus that resemble the predictions for
SrCu$_2$(BO$_3$)$_2$.  

{\it Effective S = 1/2 model:} For the interpretation of the
experimental data at low temperatures we use an effective
$S = 1/2$ Ising model on the SSL with afm exchange $J_1 > 0$ on the
square bonds and $J_2 > 0$ on diagonal bonds. In the ideal Ising limit
(zero transverse exchange) the ground state of the model is the N\'{e}el state
for $J_1 > J_2/2$ and an Ising dimer state (opposite spins on the SS
diagonals) for $J_1 < J_2/2$. Starting from the N\'{e}el ground state, an
applied field induces a 1/2 magnetization plateau for
$2J_1 - J_2/2 < \mu_B B < 2J_1 + J_2/2$ and saturation at still higher fields
with discontinuous transitions between these states. From the data on the
extent of the N\'{e}el and 1/2 plateau phases we conclude $J_2 \approx J_1$,
in agreement with the exchange paths expected from the geometric structure.
An additional transverse exchange (exchanging effective spin components 
$+\frac{1}{2}$ and $-\frac{1}{2}$) will broaden the discontinuous transitions 
in an applied field. Transverse exchange, however, will be small due to the 
strong single ion anisotropy and numerical calculations of the magnetization 
for a 16 spin SSL show that the broadening is below the experimentally observable
limit for an xy-anisotropy $\epsilon < 0.1$.  

For an understanding of the rich superstructures and of the fractional plateaus observed in the regime
intermediate between afm and ferrimagnetic order,
the Ising limit is not sufficient (the spatial period in this regime is just
doubled compared to the period of the N\'{e}el phase) and in fact
several higher order corrections may contribute to the observed
features such as additional exchange interactions between further
neighbors resulting from RKKY interactions and magnetoelastic effects
\cite{BisLPB07}.

Whereas RKKY interaction effects appear as the natural choice to
describe complex spatial structures in a rare earth compound and have, in
addition to Ising exchange, in fact been used to describe e.g.~a variety of
spatial structures in the compound CeSb \cite{Date88}, this will
require very special interaction ratios over large distances. Considering the
similarity to SrCu$_2$(BO$_3$)$_2$ this nongeneric description is not
satisfying and we find it worthwhile to consider as alternative
possibility the influence of the transverse exchange using 
the concepts applied to magnets on the
SSL before \cite{JolMG02}. These concepts in fact do provide a much more
natural approach for an understanding of the experimental findings in
TmB$_4$. In the following we thus speculate that TmB$_4$ could be considered
as equivalent to a 2D spinless fermion gas in a very strong magnetic
field. This will lead to a proposal for the understanding of plateaus and the
accompanying spatial structures in parallel to the corresponding
interpretation of the high field phenomena in the well investigated compound
SrCu$_2$(BO$_3$)$_2$ \cite{JolMG02,Seb07,MomT00}. This interpretation is based
on (i) the mapping of spins 1/2 to hard core bosons and (ii) the subsequent
mapping of hard core bosons to fermions. In 2D, step (ii) requires the
introduction of a fictitious magnetic field, the Chern-Simons field $H_{\rm
CS}$ with vector potential $\mathbf{a}(r)$ and flux $\Phi(r)$ determined by
the fermion density $\rho(r)$, given by $\nabla \times {\mathbf a} = \phi_0 \
\rho(r) {\mathbf e_z}$ and $\Phi(r) = 2 \pi \rho(r)$.  Here, $\phi_0$ is the
flux quantum and $\Phi(r)$ is the flux through a plaquette adjacent to lattice
point $r$ (choose once and for all one of the 4 plaquettes adjacent to a point
$r$ with the fermion density $\rho(r) = S^z_r+\frac{1}{2}$). Apart from this
fictitious magnetic field $H_{\rm CS}$ there is an effective fermion
hamiltonian with band width determined by the transverse exchange and
interaction determined by the Ising part of the magnetic exchange $J^z$.

If the nonlocal vector potential attached to each fermion is treated in the
average flux approximation as described in Ref.[\onlinecite{JolMG02}], the
problem is reduced to a gas of interacting spinless fermions in a strong
magnetic field characterized by an average flux related to the magnetization
$M$ of the original magnetic model
\[ \Phi = 2\pi \langle \rho(r) \rangle \ = 
       2\pi \ \left( \langle S^z \rangle + \frac{1}{2} \ \right) \ = 
       2\pi \ \left(M + \frac{1}{2} \ \right) \]
A noninteracting electron gas on the square lattice has been studied starting
from a tight binding band \cite{Hof76} as well as from Landau levels
\cite{RauWO74,ThoKNN82,Str82,SprKG97}, both approaches showing the emergence
of a fractal structure due to the periodic lattice potential, the Hofstadter
butterfly. Starting from Landau levels leads more directly to the quantum Hall
effect: Landau band splits into $p$ subbands, minigaps result which should be
observable in additional plateaus of the Hall conductance. Whereas a direct
experimental observation is not possible due to the extremely high magnetic
fields required, these minigaps may have been observed experimentally using
superlattice structures \cite{AlbSKW03}. In our case the relevant magnetic
field is not the physical field but the fictitious field from the Chern-Simons
term which produces magnetic flux of the order of a flux quantum per site. It
has been shown in Ref.[\onlinecite{JolMG02}] that the conductance plateaus
resulting from the combination of discreteness and fictitious Chern-Simons
field can be identified as some of the magnetization plateaus observed in
SrCu$_2$(BO$_3$)$_2$. This was done by using the SSL instead of the
square lattice and treating the electron interaction in mean field
approximation. The present case is similar in that the SSL applies also
for TmB$_4$, but is different in that the electron interaction is strong due
to the dominating Ising term in the magnetic hamiltonian. Thus the situation
is that of electrons on a SSL in the regime of the FQHE and the 
plateaus of interest will result from the combined effect
of the lattice (leading to minigaps) and of electron-electron interactions
(leading to the gaps of the FQHE), a combination which is expected to result
in a large variety of possible plateaus. We therefore speculate that TmB$_4$
is a magnetic analogon of the FQHE on a periodic lattice.

Our investigations have shown that the Ising like 2D SS magnet
TmB$_4$ is characterized by unusual magnetization plateaus at small fractional
values (1/7, 1/9 $\dots$) , tied to complex spatial structures. We have
speculated that a consistent description of plateaus and spatial structures
may be obtained from a mapping of this model to the FQHE in a 2D electron gas
with strong Coulomb interactions on a lattice in high field. Whereas
alternative interpretations should not be excluded at the moment, our
speculation, if true, would open up the new field of studying quantum states
similar to the FQHE in magnetic materials. We hope that this work
stimulates further investigations although, in view of the
efforts to relate microscopic numerical calculations for interacting electrons
to the FQHE \cite{HalR85}, the numerical work required may be beyond the present possibilities.

{\it Acknowledgements:} We have the pleasure the acknowledge stimulating
discussions with R. Haug, B. Lake, M. Meissner, A. Tennant and K. Totsuka. This work in
part was supported by INTAS and DAAD, by Slovak Agencies VEGA (7054) and APVT
(0166), and US Steel Kosice.

\end{document}